\documentclass[a4paper,12pt,fleqn]{article}

\usepackage[left=2.5cm,right=2.5cm,top=2.5cm,bottom=2.5cm]{geometry} % one column

\usepackage[latin1]{inputenc}
\usepackage{epsfig,graphics,graphicx,color}
\usepackage{amsmath,amssymb,amsthm,latexsym,mathrsfs}

\usepackage[pdftex,hyperfootnotes=false,pdfpagelabels,pdfstartview=FitH]{hyperref}
\usepackage[bf,font={singlespacing,small}]{caption}
\usepackage{subfig}
\usepackage{setspace}

\newcommand{\mailto}[1]{\href{mailto:#1}{#1}}

\title{\textbf{A mechanical model of the smartphone's accelerometer}}

\author{\textbf{Aurelio Agliolo Gallitto}$^{(1)}$, \textbf{Lucia Lupo}$^{(2)}$\vspace{2mm} \\
$^{(1)}$Dipartimento di Fisica e Chimica, Università di Palermo \\ via Archirafi 36, I-90123 Palermo, Italy\\
$^{(2)}$Liceo Scientifico Statale ``Galileo Galilei'' \\ via Danimarca 54, I-90146 Palermo, Italy}

\date{}

\begin{document}

\maketitle

% \begin{abstract}
%
% \end{abstract}

\begin{spacing}{1.2}
% \noindent \textbf{PACS numbers} \vspace{6pt}\\
% \begin{tabular}{ll}
%  65.20.+w & Thermal properties of liquids: heat capacity, thermal expansion, etc. \\
%  01.50.Pa & Laboratory experiments and apparatus\\
%  44.25.+f & Natural convection\\
%  01.30.la & Secondary schools
% \end{tabular}

%\keywords{}

\vfill

\noindent \textbf{Corresponding author} \vspace{6pt}\\
Prof. Aurelio Agliolo Gallitto\\
Dipartimento di Fisica e Chimica, Università di Palermo\\
via Archirafi 36, I-90123 Palermo, Italy\\
Tel: +39 091 238.91702 -- Fax: +39 091 6162461\\
E-mail: \mailto{aurelio.agliologallitto@unipa.it}
\end{spacing}

\newpage
\noindent In the recent literature, an increasing number of papers are devoted to the use of mobile media devices (smartphones, tablets, etc.) as laboratory tools for school physics experiments \cite{oprea,cabeza,kuhn,vogt}, since they are usually equipped with several sensors that can be controlled by appropriate software (apps) \cite{straw}.
Beside microphone and speaker, all the smartphones are equipped by an acceleration sensor, the accelerometer, which senses the device's orientation to set the display orientation.

The accelerometer allows one to perform a large number of quantitative experiments in mechanics \cite{vogt}, to be carried out in the classroom.
However, before introducing the use of accelerometer in lessons, it is very fruitful to understand how it works \cite{kuhn,vogt}.
Usually, one-axis acceleration sensor consists of a prof mass, mounted on springs, that can move freely in one direction.
When the device on which the accelerometer is located accelerates, it causes the mass to move by a certain distance.
The variation of the mass position, which can be measured with several different methods, provides a way to determine the acceleration of the device.

The most used method to detect the variation of the mass position is the capacitive method \cite{kuhn}.
In this case, in the device it is implemented a micro-machined capacitive sensing cell consisting of two outer fixed plates and an inner moveable plate, which deflects from its rest position whenever the system accelerates.
The three plates form two back-to-back capacitors, that is a series connection of two capacitors.
When the center plate accelerates, the distance from it to one of the fixed plates will increase whereas that to the other plate decreases by the same amount.
As the distance between the plates changes, each capacitor's value will change.
An integrated circuit measures the two capacities and extracts the acceleration data from the difference between the two values, providing an output voltage proportional to the acceleration.
The values of acceleration is obtained after the accelerometer is properly calibrated.
We would like to remark that the calibration procedure of the smartphone should be performed with the device at rest on a horizontal plane to set the values of the gravity acceleration $g = 9.806~\mathrm{m~s^{-1}}$.

To measure the three components of acceleration independently, three sensors have to be positioned orthogonally to each other that measure the acceleration components $a_x$, $a_y$ and $a_z$ of each spatial direction $x$, $y$ and $z$, respectively.

\begin{figure}[b!]
  \centering
  \includegraphics[height=0.4\textheight]{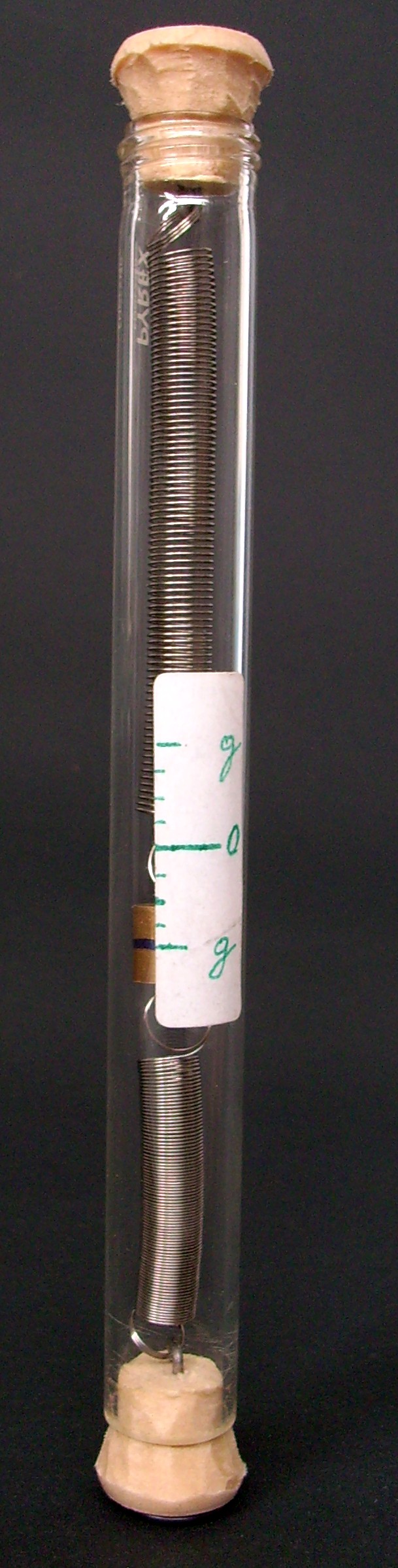} \hspace{0.5cm}
  \includegraphics[height=0.4\textheight]{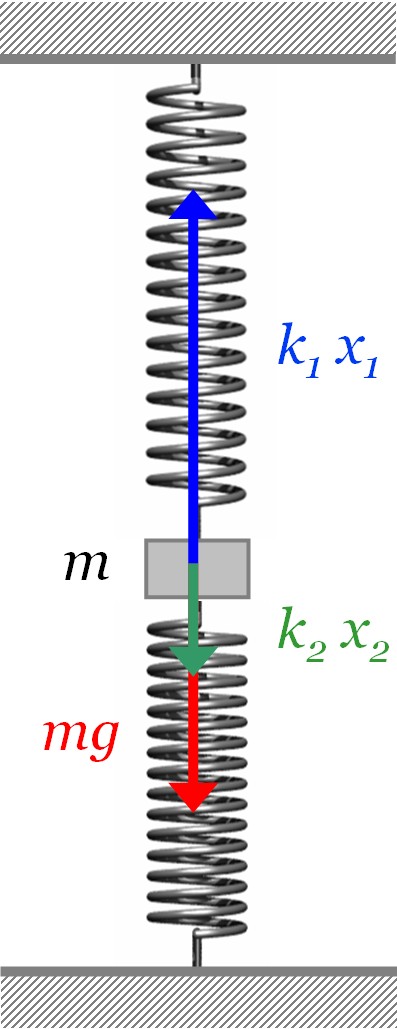}
  \caption{Mechanical model (left) and scheme (right) of the one-axis accelerometer.}
  \label{fig:modello-accelerometro}
\end{figure}

In order to allow students to better understand how the accelerometer works, we have developed a mechanical model that consists of a proof mass $m \approx 10$~g connected in series with two stretched springs having equal elastic constant $k_1 \approx k_2 \approx 5~\mathrm{N~m^{-1}}$.
In figure \ref{fig:modello-accelerometro}, it is shown the mechanical model and scheme of the accelerometer.

When the accelerometer is located horizontally, the proof mass remains in equilibrium under the action of the elastic forces of the two springs: $k_1 x_1 - k_2 x_2 = 0$, where, $k_1$ and $k_2$ are the elastic constants of the two springs; $x_1$ and $x_2$ are the correspondent elongations.
On the contrary, when the accelerometer is moved vertically with acceleration $a$, the proof mass is subjected also to the gravity force and the general motion equation is $k_1(x_1 + x) - k_2(x_2 - x) - mg = ma$, where $x$ is the additional elongation/compression of the springs.
From this equation, considering that in our case, $k_1 = k_2$ and then $x_1 = x_2$, one obtains the relation between the displacement and the acceleration of the proof mass:
\begin{equation}\label{eqn:x}
x = \frac{m}{2k}(g + a)\,.
\end{equation}
When the proof mass is at rest, $a=0$ and then $x = \frac{m}{2k}g \approx 10$~mm.
When the accelerometer is in free fall, the acceleration of the proof mass is $a = - g$ and from equation \eqref{eqn:x} one obtains $x=0$.

Usually, the devices' operative system allows one to directly display the values of acceleration on the mobile device.
However, on the web now it is possible to find several apps that allow one also to plot a graph of $a_x$, $a_y$, $a_z$ as a function of time \cite{straw}.
Therefore, dropping the smartphone onto a pillow it will give the expected graph: $a = 0$ when the devices in free fall and some quite large values when it hits the pillow.

In conclusion, we have developed a mechanical model of the smartphone's accelerometer, which can be used in classroom to allow students to better understand how the smartphone's accelerometer works.
Although smartphones are very sofisticate devices, the principle of the accelerometer can be easily understood even by students at the beginning of the study in physics.
To increase the attention of students, several physics experiments can be performed at school, as well at home, by using the smartphone as laboratory tools, a very well know device more and more diffused among students.

\end{document}